# DEVELOPING A SAAS-CLOUD INTEGRATED DEVELOPMENT ENVIRONMENT (IDE) FOR C, C++, AND JAVA

**[1]A.B. MUTIARA, [2]R. REFIANTI, [3]B.A WITONO**

[1]Prof, Faculty of Computer Science and Information Technology, Gunadarma University, Indonesia

[2]Asst.Prof, Faculty of Computer Science and Information Technology, Gunadarma University, Indonesia

[3]Alumni, Faculty of Computer Science and Information Technology, Gunadarma University, Indonesia

E-mail: [1,2]{amutiara,rina}@staff.gunadarma.ac.id, [3]banuaw@student.gunadarma.ac.id

**ABSTRACT**

Cloud era brought revolution of computerization world. People could access their data from anywhere and anytime with different devices. One of the cloud's model is Software as a Service, which capable to provide applications that run on a cloud infrastructure.An IDE (Integrated Development Environment) is the most popular tool to develop application in the network or single computer development. By installing IDE in each computer of the network could causes the lot of time and budget spending. The objective of the research is developing an efficient cloud based IDE. The IDE could compile the code which sent from client browser through SaaS IDE to the server and send it back to the client. The method that used in the research is the System Development Life-Cycle: Waterfall and Unified Model Language as system designing tool. The research successfully produced the cloud-based SaaS IDE with excellent result from several testing in local network and internet.

## 1. INTRODUCTION

Computer programming is the one of process of building applications. Starting from a source code written, tested, maintained and repaired. Source code is derived from one or more programming languages. Programming language is the bridge between human and computer. Human could interact with computer through programming language. Learning a programming language requires at least an application to compile or interpret written program to run. Integrated Development Environment (IDE) is the most popular tool in building an application [5]. Compiler or interpreter needed to be installed first before the IDE display the result of written code. Sometimes the compiler is included in the IDE. Building an application could be in a standalone computer or in a computer which connected to the network. Application development in the network is an efficient way to build an application. Application that is built in a network requires installation of an IDE for each connected computer, so that every computer should have available storage resources.

Technology development affects the birth of the cloud computing. The publication from NIST defines cloud computing as a model for enabling ubiquitous, convenient, on-demand network access to a shared pool of configurable computing resources (e.g., networks, servers, storage, applications, and services) that can be rapidly provisioned and released with minimal management effort or service provider interaction. Many applications can be accessed via the cloud, so users can access them anywhere and anytime. Cloud technology has three service models: IaaS (Infrastructure as a Service), PaaS (Platform as a Service), and SaaS (Software as a Service) [1]. SaaS provides software as a service from cloud provider, could be PaaS or any form of cloud server. The example are office application, customer relationship management, enterprise resource planning, management information systems, content management, etc. The other examples of cloud-based applications are email service and online storage. SaaS is not just an application that runs via the browser, but also desktop applications that connected to a network or the internet. Dropbox is an example of a SaaS application that runs as a desktop.

The characteristics of cloud computing changed the paradigm of application usage. Users do not need to install various types of applications, only need to install a browser on a computer connected to the internet and can use applications. This leads to minimize the use of computer storage and increase portability. Building a cloud-based applications can be through three models: private, community, public, and hybrid network. This model was chosen according to the needs of cloud application development. The scope of the research is developing the cloud-based IDE which used to develop small scale application. The provided languages are: C,C++, and Java. Those languages are in the top rank of language popularity from langpop.com.It is developed in local network and tested in local and global network or internet. The result of compiled code would tested based on five basic functions: basic input output, basic numeric operations, string operation, decision making, and loop control.

The main objectives of this research is Developing cloud-based SaaS Integrated Development Environment (IDE) and the IDE that could compile code through server and send the result to the browser.

## 2. RESEARCH METHODS

In this research there are several step of the method in order to achieve the purpose of research. The research method is the System Development Life-Cycle: Waterfall. The steps are:

a. Analysis: system analysis is the process of decomposition of a complete information system into component parts with a view to identify and evaluate problems, opportunities, barriers and needs that are expected to occur in order to proposed improvements. Knowing the general system of IDE which installed in the computer.

b. Designing: system design is the process of making the design work flow management and design of programming needed for the development of information systems. The design of the system is developed using the tools of UML (Unified Modeling Language), navigation structure, database, and user interface design.

c. Implementation: implementation is the process to complete the approved system design and test, install, start, and use the new system or systems are developed. Developing a SaaS applications using development tools like XAMPP with Bluefish, PHP programming language for web programming and MySQL for database.

d. Testing: testing activity is intended to test the cloud-based SaaS application according to the objective of the research. In this step would explain the result of functional testing, compatibility testing, browser's performance testing and the additional testing due to test the result of compiled code between desktop IDE and cloud-based IDE.

## 3. THEORY

### 3.1 Cloud Computing

Cloud computing is a model for enabling ubiquitous, convenient, on-demand network access to a shared pool of configurable computing resources (e.g., networks, servers, storage, applications, and services)that can be rapidly provisioned and released with minimal management effort or service provider interaction. This cloud model is composed of five essential characteristics, three service models, and four deployment models. Essential Characteristics of cloud [1]:

a. On-demand self-service.

A consumer can unilaterally provision computing capabilities, such as server time and network storage, as needed automatically without requiring human interaction with each service provider.

b. Broad network access.

Capabilities are available over the network and accessed through standard mechanisms that promote use by heterogeneous thin or thick client platforms (e.g., mobile phones, tablets, laptops, and workstations).

c. Resource pooling.

The provider's computing resources are pooled to serve multiple consumers using a multi-tenant model, with different physical and virtual resources dynamically assigned and reassigned according to consumer demand. There is a sense of location independence in that the customer generally has no control or knowledge over the exact location of the provided resources but may be able to specify location at a higher level of abstraction (e.g., country, state, or datacenter). Examples of resources include storage, processing, memory, and network bandwidth.

d. Rapid elasticity.

Capabilities can be elastically provisioned and released, in some cases automatically, to scale rapidly outward and inward commensurate with demand. To the consumer, the capabilities available for provisioning often appear to be unlimited and can be appropriated in any quantity at any time.

e. Measured service.

Cloud systems automatically control and optimize resource use by leveraging a metering capability1 at some level of abstraction appropriate to the type of service (e.g., storage, processing, bandwidth, and active user accounts). Resource usage can be monitored, controlled, and reported, providing transparency for both the provider and consumer of the utilized service.

Service Models of cloud [1]:

a. Software as a Service (SaaS).

The capability provided to the consumer is to use the provider's applications running on a cloud infrastructure. The applications are accessible from various client devices through either a thin client interface, such as a web browser (e.g., web-based email), or a program interface. The consumer does not manage or control the underlying cloud infrastructure including network, servers, operating systems, storage, or even individual application capabilities, with the possible exception of limited user- specific application configuration settings.

b. Platform as a Service (PaaS).

The capability provided to the consumer is to deploy onto the cloud infrastructure consumer-created or acquired applications created using programming languages, libraries, services, and tools supported by the provider The consumer does not manage or control the underlying cloud infrastructure including network, servers, operating systems, or storage, but has control over the deployed applications and possibly configuration settings for the application-hosting environment.

c. Infrastructure as a Service (IaaS).

The capability provided to the consumer is to provision processing, storage, networks, and other fundamental computing resources where the consumer is able to deploy and run arbitrary software, which can include operating systems and applications. The consumer does not manage or control the underlying cloud infrastructure but has control over operating systems, storage, and deployed applications; and possibly limited control of select networking components (e.g., host firewalls).

**3.2 Integrated Development Environment**

An Integrated Development Environment (IDE) is an application that facilitates application development. In general, an IDE is a graphical user interface (GUI)-based workbench designed to aid a developer in building software applications with an integrated environment combined with all the required tools at hand. Most common features, such as debugging, version control and data structure browsing, help a developer quickly execute actions without switching to other applications. Thus, it helps maximize productivity by providing similar user interfaces (UI) for related components and reduces the time taken to learn the language. An IDE supports single or multiple languages [2]. An Integrated Development Environment (IDE) brings all of the programmers tools into one convenient place. There was a time when programmers had to edit files, save the files out, run the compiler, then the linker, build the application then run it through a debugger. Today's IDEs bring editor, compiler, linker and debugger into one place along with project management tools to increase programmer productivity [3,5,6,7,8].

**3.3 Shell**

The shell provides us with an interface to the UNIX system. It gathers input from you and executes programs based on that input. When a program finishes executing, it displays that program's output. A shell is an environment in which we can run our commands, programs, and shell scripts. There are different flavors of shells, just as there are different flavors of operating systems. Each flavor of shell has its own set of recognized commands and functions. The prompt, $, which is called command prompt, is issued by the shell. While the prompt is displayed, we can type a command. The shell reads your input after you press Enter. It determines the command you want executed by looking at the first word of your input. A word is an unbroken set of characters. Spaces and tabs separate words [4].

In UNIX there are two major types of shells [4]:

• The Bourne shell. If you are using a Bourne-type shell, the default prompt is the $ character.

• The C shell. If you are using a C-type shell, the default prompt is the % character.

There are again various subcategories for Bourne Shell which are listed as follows:

• Bourne shell (sh)

• Korn shell (ksh)

• Bourne Again shell (bash)

• POSIX shell (sh)

The different C-type shells follow:

• C shell (csh)

• TENEX/TOPS C shell (tcsh)

The original UNIX shell was written in the mid-1970s by Stephen R. Bourne while he was at AT&T Bell Labs in New Jersey. The Bourne shell was the first shell to appear on UNIX systems, thus it is referred to as "the shell". The Bourne shell is usually installed as /bin/sh on most versions of UNIX. For this reason, it is the shell of choice for writing scripts to use on several different versions of UNIX. Shell can be used for compile any programming language in any platform [4].

## 4. RESULTS AND DISCUSSION

### 4.1 Analysis

Generally, the Integrated Development Environment (IDE) must be installed first along the compiler before it ready for use [5,7,8]. For example, when users would developing Java application using Netbeans, they must install Java Development Kit as package to compile and build the code. It doesn't matter if user really need to compile the large scale application using IDE such like Netbeans. But if the user who just want to learn how to code, it really annoying to install and provide some space to IDE. Installing IDE in local network also require long process. For example, in the University which has Computer Science diploma must have computer laboratory. Laboratory needs to install IDE to run the specific programming code in programming class. Installing IDE in different machine one by one even in the local network like Figure 2 is really need a lot of time. The other problem also come when the laboratory assistant need to check the task file of IDE in several computer one by one.

Cloud-based IDE brings potential business in cloud platform, for example cloud provider may build cloud-based SaaS IDE to run the business. By providing their storage and IDE as a service, user can write and run code anywhere even without their own device. Another opportunity for this theory is collaborative working for developer team. Many project needed to build application by collaborative way, so person in the team could work together in separate place but still connected to the network/internet. This research explain different approach for learning programming code or developing small scale application in network like internet, without installing any IDE or packages for user computer.

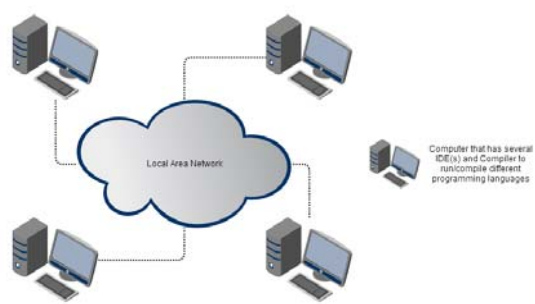

Figure 2. IDE in Local Network Architecture

### 4.2 Defining the new model

The new system design offer many efficiency without any major changes in any cloud network. Local cloud architecture could contains of several components. In the cloud server could containing infrastructure, platform, service, or storage. In this research would bring explanation about cloud network which provide software as a service (SaaS). SaaS means that user only access the software and storage in the cloud architecture, without knowing what inside the cloud server. The system designed is a system that adopts cloud architecture in general. A server connected to a network/internet where other computers in the network/internet be able to connect and use the IDE. Global network architecture shown in Figure 3.

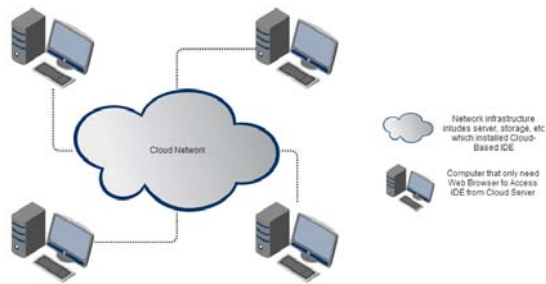

Figure 3. Cloud-Based IDE in Cloud Architecture

In cloud server, the IDE has been installed along with compiler packages. Means that user as client no need to install several packages to compile code. Using few server configurations that will be explain later in the next chapter. For local network server runs on pre-installed operating system and connected to wireless router. Wireless router also has configured for being an AP (Access Point) device. When others computer connected to the AP device, the computers can use the IDE along with compiler in it. For internet architecture, the IDE only need to installed in the cloud server via SSH or other remote model. Client only need to install browser or maybe using preinstalled browser. Client would access to the address that could be IP or hostname. Browser would display the IDE from server trough HTTP protocol. Interface will be supported by JavaScript and CSS customization. The feature of the IDE would depend on SaaS characteristics.

### 4.3 Result

#### 4.3.1 Login Page

Login page is getting information (username and password) from user as validation before user enter the cloud-based IDE, as shown in Figure 4.

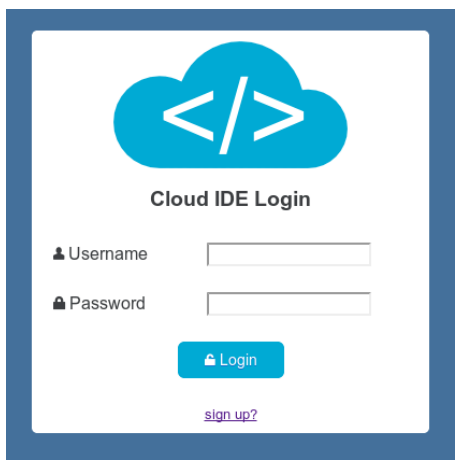

Figure 4. Login page for administrator and user

#### 4.3.2 Administrator Dashboard

This page just like another welcome page for administrator. Showing some utility functions to monitor the application and navigation to other pages as shown in Figure 5. The administrator could see the active user, registered user, compiled file, directory space, and last active session in this page.

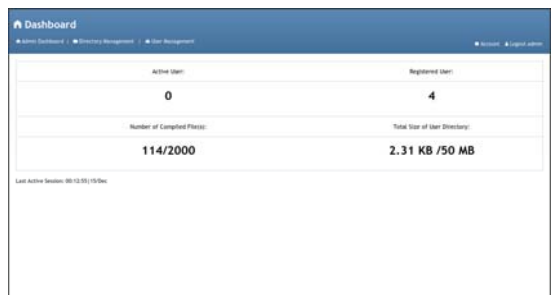

Figure 5. Dashboard for administrator

#### 4.3.3 Directory management

The page shows the file manager of all user directory. By clicking right-mouse button will display a context menu as shown in Figure 6. Context menu shows the operation of file or folder from all registered user. Set warming limitation section is used for control the disk space and compiled limitation of the application. It's useful if the application is run on cloud server that provide by the provider (not build by own architecture).

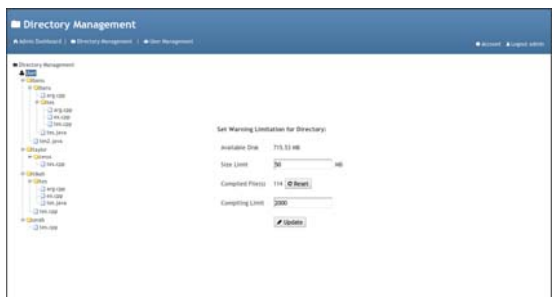

Figure 6. Directory management page

#### 4.3.4 User management

The page displays several information to manage user data, and monitor their status as shown in Figure 7. Showing the all registered users, their disk limitation, and current network status. From this page, administrator could edit their information by clicking manage button.

Figure 7. User management page

#### 4.3.5 IDE page

The core page of the IDE that was built for the research. Contains of menu bar which have several function for user such like run button, file manager which for manipulate user file or folder, code editor for write the code in certain languages and the last is output bar, which used for debugging or displaying result from compiled code as shown in Figure 8. The snippet code from IDE layout can be seen in appendix, all line codes from this page are written in HTML and Javascript Language.

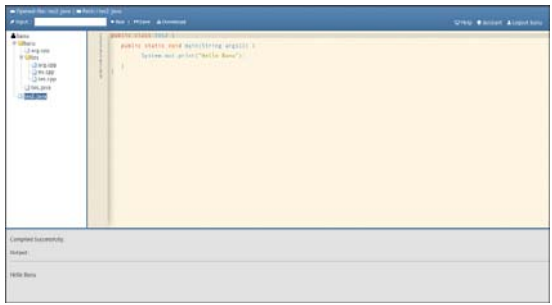

Figure 8. IDE page

The input text value is filled for argument that pre-requested by user by clicking run button, user will receive result in few seconds depends on the code and server's performance. The process can be seen in sequential diagram in Figure 9.

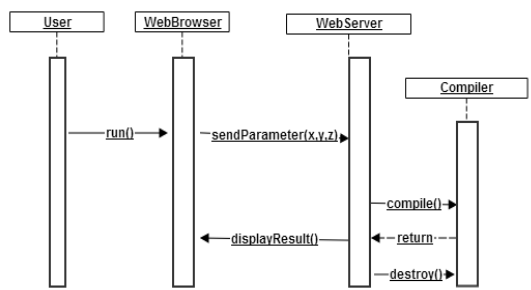

Figure 9. Sequence diagram for compiling process

It start from getting user identity and database configuration. Then it check the limitation from administrator have set. If the limit larger than the current compiling number, the process will continue. This condition start from defining certain variable that have values same as requested data from user. The extension of file that would be compiled is getting by program. After that, the code will be saved in specific directory as temporary and it will be opened and execute by shell in server through `shell_exec()` function in php file. The result of the compiled code will be displayed as html code in output bar. If the result is not displayed in shell, it won't display and it must be download to see the result. The example for this case is graphical application that produce from compiling code. For the download button the process is quietly same like run button, but for download button, the result of compiled will be downloaded through the user browser after compiling process.

### 4.4 Testing

There are several type of cloud testing: stress test, load & performance test, functional test, compatibility test, browser performance test, and latency test. In this section would explain the result of functional testing, compatibility testing, and browser's performance testing. The additional testing is given due to test the result of compiled code between desktop IDE and cloud-based IDE. The success percentage of the functional and similarity testing are according to the following equation:

$$SP = \frac{\sum sample\ were\ detected}{\sum sample\ tested} x 100\%$$

SP= Success Percentage or Similarity

#### 4.4.1 Functional Testing

This testing based on test case that had been made. The IDE was installed in local server and cloud VPS. Functional testing result divided by two tables, table 1 for administrator, and table 2 for user.

Table 1: Functional Testing for Administrator

| No | Function | Input | Expected Result | Output | Status |
|---|---|---|---|---|---|
| 1 | log in | Valid sign in | Displaying Dashboard | Displaying dashboard | Success |
| 2 | Directory management menu | Opening the menu | Displaying directory management | Displaying directory management | Success |
| 3 | Folder creation | Create folder | Appear new folder | Appear new folder | Success |
| 4 | File creation | Create file | Appear new file | Appear new file | Success |
| 5 | File renaming | Rename file | File name changed | File name changed | Success |
| 6 | File downloading | Download file | File Downloaded | File Downloaded | Success |
| 7 | File deletion | Delete file | File deleted | File deleted | Success |
| 8 | Folder deletion | Delete folder | Folder deleted | Folder deleted | Success |

| | | | | | | | | | | | | |
|---|---|---|---|---|---|---|---|---|---|---|---|---|
| | | | | | | | | button | | | | |
| 9 | Warning limitation update | Update warning limit | Updated limitation | Updated limitation | Success | | 12 | Help page | Click help button | Displaying help | Displaying help | Success |
| 10 | User management menu | Open the menu | Displaying the menu | Displaying the menu | Success | | 13 | Account menu | Select account | Displaying user info | Displaying user info | Success |
| 11 | User info selection | Select a user | Selected user | Selected user | Success | | 14 | Password update | Update password | Updated password | Updated password | Success |
| 12 | User info update | Update user | Updated user info | Updated user info | Success | | 15 | Log out | Sign out | Displaying login page | Displaying login page | Success |
| 13 | Account menu | Select account | Displaying user info | Displaying user info | Success |
| 14 | Password update | Update password | Updated password | Updated password | Success |
| 15 | Log out | Sign out | Displaying login page | Displaying login page | Success |

The success percentage of administrator functional testing:

$$SP = \frac{15}{15} x 100\%$$

$$SP = 100\%$$

Table 2: Functional Testing for User

| No | Function | Input | Expected Result | Output | Status |
|---|---|---|---|---|---|
| 1 | Registration | Insert valid data | Displaying login page | Displaying login page | Success |
| 2 | Log in | Sign in | Displaying IDE | Displaying IDE | Success |
| 3 | Folder creation | Create folder | Appear new folder | Appear new folder | Success |
| 4 | File creation | Create file | Appear new file | Appear new file | Success |
| 5 | File renaming | Rename file | File name changed | File name changed | Success |
| 6 | File downloading | Download file | File Downloaded | File Downloaded | Success |
| 7 | File deletion | Delete file | File deleted | File deleted | Success |
| 8 | Folder deletion | Delete folder | Folder deleted | Folder deleted | Success |
| 9 | Code writing | Write valid code | Displaying syntax highlighting | Displaying syntax highlighting | Success |
| 10 | Code compiling | Click run button | Displaying result | Displaying result | Success |
| 11 | File saving | Click save | File saved | File saved | Success |

The success percentage of user functional testing:

$$SP = \frac{15}{15} x 100\%$$

$$SP = 100\%$$

### 4.4.2 Compatibility Testing

This section describes the result from compatibility several platform and browser, during running the IDE as shown in Table 3. The platform are Windows, Linux (PC), and Android. Each of them has one or more browser tested. The all devices are connected in the same network with server.

Table 3: Compatibility Testing

| Platform | Windows | | | Linux | | Android Browser |
|---|---|---|---|---|---|---|
| Browser | Chrome 31 | Firefox 22 | IE 10 | Chrome 31 | Firefox 22 | Default 2.3 |
| Status | Run Full Function | Run Full Function | Run Full Function | Run Full Function | Run Full Function | Run Partial Function |
| | | | | | | Unable run context menu |

### 4.4.3 Browser Performance Testing

The browser performance test describes the result of performance from three popular browsers. The IDE was installed in cloud VPS. The browsers are installed in Windows platform and connected to the server (Ubuntu 12.04). As shown in Table 4 there are five times of trial to get average result of load time, working css, and working function. Load time performance is analyzed by pingdom tool to display time in milli second.

Table 4: Browser Performance Testing

| Number of Trial | Load time (MS) | | | Working CSS and Running Function | | |
|---|---|---|---|---|---|---|
| | Chrome 31 | Firefox 22 | IE 10 | Chrome 31 | Firefox 22 | IE 10 |
| 1 | 230 | 461 | 255 | All | All | Glitch CSS in file manager |
| 2 | 260 | 270 | 268 | | | |
| 3 | 249 | 214 | 225 | | | |
| 4 | 226 | 216 | 226 | | | |
| 5 | 268 | 145 | 262 | | | |
| Average | 247 | 261 | 249 | | | |

According to the result, there is glitch in the css of the file management section. It caused by incompatibilty of the browser (IE 10) but it doesn't affect seriously with the function of IDE.

**4.4.4 Result Similarity Testing**

Result comparison test describes the comparison between result of compiled code in desktop IDE and cloud-based IDE. The testing uses same version of compiler packages. Accessing the IDE which installed in local cloud network. The result obtain 100% value according to the testing equation. As shown in Table 5, the cloud IDE has similar result with desktop IDE, which has five kind of activity.

Table 5: Result Comparison Testing

| Activity | Similarity of Result | | |
|---|---|---|---|
| | Java | C++ | C |
| Basic I/O | yes | yes | yes |
| Basic Numeric Operation | yes | yes | yes |
| String Operation | yes | yes | yes |
| Decision Making | yes | yes | yes |
| Loop Control | yes | yes | yes |

The success percentage of similarity is:

$$SP = \frac{15}{15}x100\%$$

$$SP = 100\%$$

## 5. CONLUSION REMARKS

**5.1 Conclusion**

In this research successfully produced the cloud-based SaaS IDE which can be used to develop C, C++, and Java. The cloud-based IDE could reduce installation time because the centralized technology of cloud. It also only need tiny client (small size of disk usage) to run IDE as client-based. Functional testing result indicated the success of all functions in the IDE start from log in the application until the end of session (log out). The compatibility testing result shown that the IDE could run well in four popular browsers, on three different platform. Based on compatibility test, the android browser unable to run context menu function in order to manage the file, it's typically problem cause of the phone browser doesn't support for right-click command. Browser performance testing from VPS shown that the IDE could be better run on Chrome. By using chrome, user could access the IDE faster than other browsers, and also with all working functions and CSS. According to comparison testing result, all the basic activity of programming (basic I/O, numeric operations, string operations, decision making, and loop control)are running with similar result as other IDEs.

**5.2 Future Works**

In Cloud-based SaaS IDE that had been built before, user could not see the result of the compiled code that use the image processing. For java it can be use Java applet to display the compiled code which have image processing in it. Note also the security of user data, because in building applications on cloud architectures, security is also a top priority. Although the function of compiling in IDE still running well in any portable device, the functions and UI must be improved for tablet or smartphone's usage. It is better to increase the performance on the server if the registered user has reach high number. Renting the high performance cloud server or build strong one, could be the solution. The shell execution is the key of compilation function, then the architecture on the cloud server might be better to use virtual machines or other virtualization methods. The improvement of virtual machine security for avoid spam or treacherous compiled code like virus application that injected to the server.